\documentstyle[aps,multicol,epsf,epsfig]{revtex}
\newcommand \ga{\raisebox{-.5ex}{$\stackrel{>}{\sim}$}}
\newcommand \la{\raisebox{-.5ex}{$\stackrel{<}{\sim}$}}
\begin{document}
\draft

\title{PHASE TRANSITIONS IN NEUTRON STARS AND MAXIMUM MASSES}
\author{H.~HEISELBERG} 

\address{Nordita, Blegdamsvej 17, DK-2100 Copenhagen \O, Denmark}

\author{M.~HJORTH-JENSEN} 

\address{Department of Physics, University of Oslo, N-0316 Oslo, Norway}

\maketitle
\begin{abstract}

Using the most recent realistic effective interactions for nuclear
matter with a smooth extrapolation to high densities including
causality, we constrain the equation of state and calculate maximum
masses of rotating neutron stars. First and second order phase
transitions to, e.g., quark matter at high densities are included.  If
neutron star masses of $\sim 2.3M_\odot$ from quasi-periodic
oscillations in low mass X-ray binaries are confirmed, a soft equation
of state as well as strong phase transitions can be excluded in
neutron star cores.

\end{abstract}
\begin{multicols}{2}
\section{INTRODUCTION}

The best determined neutron star masses are found in binary pulsars
and all lie in the range $1.35\pm 0.04 M_\odot$ (see Thorsett and
Chakrabarty 1999) except for the nonrelativistic pulsar PSR J1012+5307
of mass\footnote{95\% conf. limits or $\sim2\sigma$}
$M=(2.1\pm 0.8)M_\odot$ (van Paradijs 1998). Several X-ray binary
masses have been measured of which the heaviest are 
Vela X-1 with $M=(1.9\pm 0.2)M_\odot$ (Barziv et al., 1999)
and Cygnus X-2 with
$M=(1.8\pm 0.4)M_\odot$ (Orosz \& Kuulkers 1999).  The recent
discovery of high-frequency brightness oscillations in low-mass X-ray
binaries provides a promising new method for determining masses and
radii of neutron stars (see Miller, Lamb, \& Psaltis 1998). The
kilohertz quasi-periodic oscillations (QPO) occur in pairs
and are most likely the orbital
frequencies $\nu_{QPO}=(1/2\pi)\sqrt{GM/R_{orb}^3}$ of accreting matter
in Keplerian orbits around neutron stars of mass $M$ and its beat
frequency with the neutron star spin, $\nu_{QPO}-\nu_s$.  According to
Zhang, Strohmayer, \& Swank 1997, Kaaret, Ford, \& Chen (1997) the accretion
can for a few QPO's be tracked to its innermost stable orbit,
$R_{ms}=6GM/c^2$. For slowly rotating stars the resulting mass is
$M\simeq2.2M_\odot({\mathrm{kHz}}/\nu_{QPO})$.  For example, the
maximum frequency of 1060 Hz upper QPO observed in 4U 1820-30 gives $M\simeq
2.25M_\odot$ after correcting for the $\nu_s\simeq275$ Hz neutron star
rotation frequency.  If the maximum QPO frequencies of 4U 1608-52
($\nu_{QPO}=1125$ Hz) and 4U 1636-536 ($\nu_{QPO}=1228$ Hz) also
correspond to innermost stable orbits the corresponding masses are
$2.1M_\odot$ and $1.9M_\odot$.  Such large masses severely restrict
the equation of state (EoS) for dense matter as addressed in the
following.

Recent models for the nucleon-nucleon interaction have reduced the
uncertainty in the nuclear EoS allowing for more
reliable calculations of neutron star properties, see Akmal,
Pandharipande, \& Ravenhall (1998) and Engvik et al.~(1997).  Likewise,
recent realistic effective interactions for nuclear matter obeying
causality at high densities, constrain the EoS severely
and thus also the maximum masses of neutron stars, see Akmal,
Pandharipande, \& Ravenhall (1998) and Kalogera \& Baym (1996). We will
here elaborate on these analyses by incorporating causality smoothly
in the EoS for nuclear matter and allow for first and second order
phase transitions to, e.g., quark matter. Finally, results are
compared with observed neutron star masses and concluding remarks
are made.

\section{THE NUCLEAR EQUATION OF STATE}

For the discussion of the gross properties of neutron stars we will
use the optimal EoS of Akmal, Pandharipande, \& Ravenhall (1998) 
(specifically the Argonne $V18 + \delta v +$ UIX$^*$ model- hereafter
APR98), which is based on the most recent models
for the nucleon-nucleon interaction
with the inclusion of a parametrized
three-body force and relativistic boost corrections. The EoS for
nuclear matter is thus known to some accuracy for densities up to a
few times nuclear saturation density $n_0=0.16$ fm$^{-3}$.  We
parametrize the APR98 EoS by a simple form for the compressional and
symmetry energies that gives a good fit
around nuclear saturation densities and smoothly incorporates
causality at high densities such that the sound speed approaches the
speed of light.
This requires that the compressional part of the
energy per nucleon is quadratic
in nuclear density with a minimum at saturation but linear at high densities
\begin{eqnarray}
    {\cal E} &=& E_{comp}(n) + S(n)(1-2x)^2 \nonumber\\
   &=& {\cal E}_0 u\frac{u-2-s}{1+s u} +S_0 u^\gamma (1-2x)^2.   
    \label{eq:EA} 
\end{eqnarray}
Here, $n=n_p+n_n$ is the total baryon density, $x=n_p/n$ the
proton fraction and $u=n/n_0$ is the ratio of the baryon density to
nuclear saturation density. The compressional term is in Eq.\
(\ref{eq:EA}) parametrized by a simple form which reproduces the
saturation density and the binding energy per
nucleon ${\cal E}_0=15.8$MeV at $n_0$ of APR98. The ``softness''
parameter $s\simeq 0.2$, which gave the best fit to the data
of APR98 (see Heiselberg \& Hjorth-Jensen 1999)  
is determined by fitting the energy per
nucleon of APR98 up to densities of $n\sim 4n_0$.
For the symmetry energy
term we obtain $S_0=32$ MeV and $\gamma=0.6$ for the best fit. The
proton fraction is given by $\beta$-equilibrium at a given density.

The one unknown parameter $s$ 
expresses the uncertainty in the EoS at high
density and we shall vary this parameter within the allowed limits in
the following with and without phase transitions to calculate mass,
radius and density relations for neutron stars.
The ``softness'' 
parameter $s$ is related to the incompressibility
of nuclear matter as $K_0=18{\cal E}_0/(1+s)\simeq 200$MeV. It 
agrees with the poorly known experimental value
(Blaizot, Berger, Decharge, \& Girod 1995), 
$K_0\simeq 180-250$MeV which does not restrict it
very well.  From $(v_s/c)^2=\partial P/\partial (n\cal{E})$, where $P$ is the
pressure, and the EoS  of Eq.\ (\ref{eq:EA}),
the causality condition $v_s\le c$ requires 
\begin{equation}
      s \ga \sqrt{\frac{{\cal E}_0}{m_n}} \simeq 0.13 \,,\label{causal}
\end{equation}
where $m_n$ is the mass of the nucleon.
With this condition we have a causal EoS that reproduces the
data of APR98 at densities up to $0.6\sim 0.7$ fm$^{-3}$. 
In contrast, the EoS of APR98 becomes
superluminal at $n\approx 1.1$ fm$^{-3}$.  For larger $s$ values
the EoS is softer which eventually leads to smaller maximum masses of
neutron stars. The observed $M\simeq 1.4M_\odot$ in binary pulsars
restricts $s$ to be less than $0.4-0.5$ depending on rotation
as shown in calculations of neutron stars
below. 

\begin{figure}[htp]\begin{center}
\input{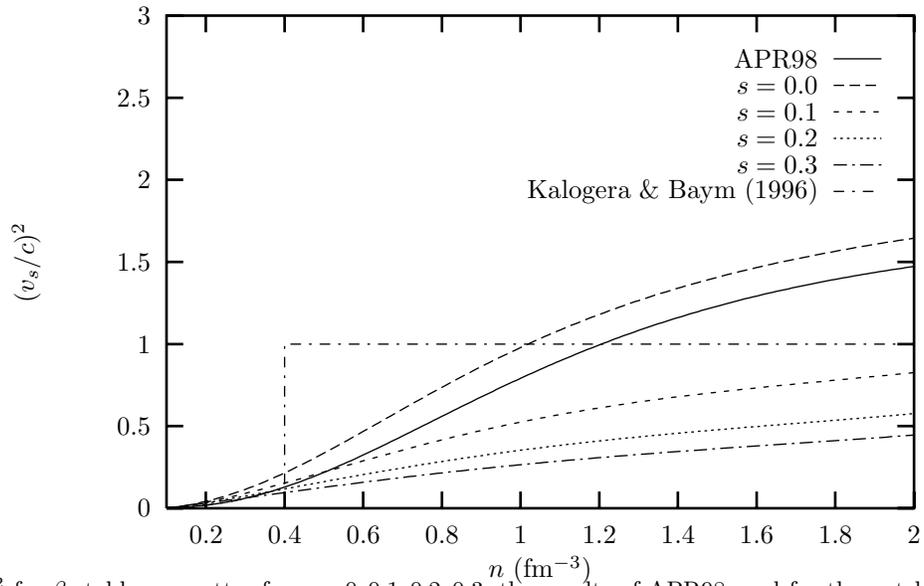}
      \caption{$(v_s/c)^2$ for $\beta$-stable $pn$-matter 
                 for $s=0,0.1,0.2,0.3$, 
                the results of APR98, 
                and for the 
                patched EoS of Kalogera \& Baym (1996) which shows 
                a discontinuous $(v_s/c)^2$. }
       \label{fig1}
\end{center}\end{figure}

In Fig.\ \ref{fig1} we plot the sound speed $(v_s/c)^2$ for
various values of $s$ and that resulting from the microscopic
calculation of APR98 for $\beta$-stable $pn$-matter.  The form of
Eq.~(\ref{eq:EA}), with the inclusion of the parameter $s$, provides
a smooth extrapolation from small to large densities
such that the sound speed $v_s$ approaches the
speed of light. For $s=0.0$ ($s=0.1$) the EoS
becomes superluminal at densities of the order of 1 (6) fm$^{-3}$.

The sound speed of Kalogera \& Baym (1996) is also plotted in Fig.\
\ref{fig1}. It jumps discontinuously to the speed of light at a
chosen density. With this prescription they were able to obtain an
optimum upper bound for neutron star masses and obey causality.  This
prescription was also employed by APR98, see Rhoades \& Ruffini (1974)
for further details.  The EoS is thus discontinuously stiffened by
taking $v_s=c$ at densities above a certain value $n_c$ which,
however, is lower than $n_{s}=5n_0$ where their nuclear EoS becomes
superluminal. This approach stiffens the nuclear EoS for densities
$n_c<n<n_s$ but softens it at higher densities. Their resulting
maximum masses lie in the range $2.2M_\odot\la M\la 2.9M_\odot$.  Our
approach however, incorporates causality by reducing the sound speed
smoothly towards the speed of light at high densities. Therefore our
approach will not yield an absolute upper bound on the maximum mass
of a neutron star
but gives reasonable estimates based on modern EoS around nuclear matter
densities, causality constraints at high densities and a smooth
extrapolation between these two limits (see Fig. 1).

\section{PHASE TRANSITIONS}

 The physical state of matter in the interiors of neutron stars at
densities above a few times normal nuclear matter densities is
essentially unknown and many first and second order phase transitions
have been speculated upon. We will specifically study the hadron to
quark matter transition at high densities, but note that other
transitions as, e.g., kaon and/or pion condensation or the presence
of other baryons like hyperons also
soften the EoS and thus further aggravate the resulting reduction in
maximum masses. Hyperons appear at densities typically of the order
$2 n_0$ and result in a considerable softening of the EoS, see
e.g., Balberg, Lichenstadt, \& Cook (1998). Typically, most equations 
of state with
hyperons yield masses around $1.4-1.6 M_{\odot}$.   
Here however, in order to focus on  the role played by phase transitions
in neutron star matter, we will assume that a phase transition
from nucleonic to quark matter takes place at a certain density.
We will for simplicity employ the bag model in our
actual studies of quark phases and neutron star properties.  In
the bag model the quarks are assumed to be confined to
a finite region of space, the so-called 'bag', by a vacuum pressure
$B$.  Adding the
Fermi pressure and interactions computed to order $\alpha_s=g^2/4\pi$,
where $g$ is the QCD coupling constant, the total pressure
for 3 massless quarks of flavor $f=u,d,s$, is (see Kapusta (1988), 
\begin{equation}
    P=\frac{3\mu_f^4}{4\pi^2}(1-\frac{2}{\pi}\alpha_s) -B +P_e+P_\mu \,,
     \label{pquark}
\end{equation}
where $P_{e,\mu}$ are the electron and muon pressure, e.g., 
$P_e=\mu_e^4/12\pi^2$.
A Fermi gas of quarks of flavor {\em i} has density $n_i =
k_{Fi}^3/\pi^2$, due to the three color states. 
The value of the bag constant {\em B} is poorly
known, and we present results using two representative values,
$B=150$ MeVfm$^{-3}$ and $B=200$ MeVfm$^{-3}$.
We take $\alpha_s=0.4$. However, similar results can be obtained with
smaller $\alpha_s$ and larger $B$  (Madsen 1998).

 The quark and nuclear matter mixed phase described in Glendenning
(1992) has continuous pressures and densities due to the general Gibbs
criteria for two-component systems. There are no first order phase
transitions but at most two second order phase transitions. Namely, at
a lower density, where quark matter first appears in nuclear matter,
and at a very high density (if gravitationally stable), where all
nucleons are finally dissolved into quark matter. This mixed phase
does, however, not include local surface and Coulomb energies of the
quark and nuclear matter structures. If the interface tension between
quark and nuclear matter is too large, the mixed phase is not favored
energetically due to surface and Coulomb energies associated with
forming these structures (Heiselberg, Pethick, \& Staubo 1993). The
neutron star will then have a core of pure quark matter with a mantle
of nuclear matter surrounding it and the two phases are coexisting by
a first order phase transition or Maxwell construction, see Fig. 2.
For a small or moderate interface tension the quarks are confined in
droplet, rod- and plate-like structures as found in the inner crust of
neutron stars (Lorenz, Ravenhall, \& Pethick 1993).

\section{NEUTRON STAR PROPERTIES}

In order to obtain the mass and radius of a neutron star, we have solved the
Tolman-Oppenheimer-Volkov equation with and without rotational
corrections following the approach of Hartle (1967).  The equations of
state employed are given by the $pn$-matter EoS with $s =0.13,
0.2, 0.3, 0.4$ with nucleonic degrees of freedon only. In addition we
have selected two representative values for the bag-model parameter
$B$, namely 150 and 200 MeVfm$^{-3}$ for our discussion on eventual phase
transitions. The quark phase is linked with our $pn$-matter EoS from
Eq.\ (\ref{eq:EA}) with $s=0.2$ through either a mixed phase
construction or a Maxwell construction, see Heiselberg and
Hjorth-Jensen (1999) for further details.  For $B=150$ MeVfm$^{-3}$, the
mixed phase begins at 0.51 fm$^{-3}$ and the pure quark matter phase
begins at $1.89$ fm$^{-3}$.  Finally, for $B=200$ MeVfm$^{-3}$, the mixed
phase starts at $0.72$ fm$^{-3}$ while the pure quark phase starts
at $2.11$ fm$^{-3}$.  In case of a Maxwell construction, in order to
link the $pn$ and the quark matter EoS, we obtain for $B=150$ MeVfm$^{-3}$
that the pure $pn$ phase ends at $0.92$ fm$^{-3}$ and that the
pure quark phase starts at $1.215$ fm$^{-3}$, while the corresponding
numbers for $B=200$ MeVfm$^{-3}$ are $1.04$ and $1.57$
fm$^{-3}$.

\begin{figure}[htp]
\setlength{\unitlength}{0.1bp}
\special{!
/gnudict 40 dict def
gnudict begin
/Color false def
/Solid false def
/gnulinewidth 5.000 def
/vshift -33 def
/dl {10 mul} def
/hpt 31.5 def
/vpt 31.5 def
/M {moveto} bind def
/L {lineto} bind def
/R {rmoveto} bind def
/V {rlineto} bind def
/vpt2 vpt 2 mul def
/hpt2 hpt 2 mul def
/Lshow { currentpoint stroke M
  0 vshift R show } def
/Rshow { currentpoint stroke M
  dup stringwidth pop neg vshift R show } def
/Cshow { currentpoint stroke M
  dup stringwidth pop -2 div vshift R show } def
/DL { Color {setrgbcolor Solid {pop []} if 0 setdash }
 {pop pop pop Solid {pop []} if 0 setdash} ifelse } def
/BL { stroke gnulinewidth 2 mul setlinewidth } def
/AL { stroke gnulinewidth 2 div setlinewidth } def
/PL { stroke gnulinewidth setlinewidth } def
/LTb { BL [] 0 0 0 DL } def
/LTa { AL [1 dl 2 dl] 0 setdash 0 0 0 setrgbcolor } def
/LT0 { PL [] 0 1 0 DL } def
/LT1 { PL [4 dl 2 dl] 0 0 1 DL } def
/LT2 { PL [2 dl 3 dl] 1 0 0 DL } def
/LT3 { PL [1 dl 1.5 dl] 1 0 1 DL } def
/LT4 { PL [5 dl 2 dl 1 dl 2 dl] 0 1 1 DL } def
/LT5 { PL [4 dl 3 dl 1 dl 3 dl] 1 1 0 DL } def
/LT6 { PL [2 dl 2 dl 2 dl 4 dl] 0 0 0 DL } def
/LT7 { PL [2 dl 2 dl 2 dl 2 dl 2 dl 4 dl] 1 0.3 0 DL } def
/LT8 { PL [2 dl 2 dl 2 dl 2 dl 2 dl 2 dl 2 dl 4 dl] 0.5 0.5 0.5 DL } def
/P { stroke [] 0 setdash
  currentlinewidth 2 div sub M
  0 currentlinewidth V stroke } def
/D { stroke [] 0 setdash 2 copy vpt add M
  hpt neg vpt neg V hpt vpt neg V
  hpt vpt V hpt neg vpt V closepath stroke
  P } def
/A { stroke [] 0 setdash vpt sub M 0 vpt2 V
  currentpoint stroke M
  hpt neg vpt neg R hpt2 0 V stroke
  } def
/B { stroke [] 0 setdash 2 copy exch hpt sub exch vpt add M
  0 vpt2 neg V hpt2 0 V 0 vpt2 V
  hpt2 neg 0 V closepath stroke
  P } def
/C { stroke [] 0 setdash exch hpt sub exch vpt add M
  hpt2 vpt2 neg V currentpoint stroke M
  hpt2 neg 0 R hpt2 vpt2 V stroke } def
/T { stroke [] 0 setdash 2 copy vpt 1.12 mul add M
  hpt neg vpt -1.62 mul V
  hpt 2 mul 0 V
  hpt neg vpt 1.62 mul V closepath stroke
  P  } def
/S { 2 copy A C} def
end
}
\begin{picture}(3600,2160)(0,0)
\special{"
gnudict begin
gsave
50 50 translate
0.100 0.100 scale
0 setgray
/Helvetica findfont 100 scalefont setfont
newpath
-500.000000 -500.000000 translate
LTa
LTb
600 251 M
63 0 V
2754 0 R
-63 0 V
600 595 M
63 0 V
2754 0 R
-63 0 V
600 939 M
63 0 V
2754 0 R
-63 0 V
600 1283 M
63 0 V
2754 0 R
-63 0 V
600 1627 M
63 0 V
2754 0 R
-63 0 V
600 1971 M
63 0 V
2754 0 R
-63 0 V
788 251 M
0 63 V
0 1795 R
0 -63 V
1163 251 M
0 63 V
0 1795 R
0 -63 V
1539 251 M
0 63 V
0 1795 R
0 -63 V
1915 251 M
0 63 V
0 1795 R
0 -63 V
2290 251 M
0 63 V
0 1795 R
0 -63 V
2666 251 M
0 63 V
0 1795 R
0 -63 V
3041 251 M
0 63 V
0 1795 R
0 -63 V
3417 251 M
0 63 V
0 1795 R
0 -63 V
600 251 M
2817 0 V
0 1858 V
-2817 0 V
600 251 L
LT0
3114 1946 M
180 0 V
656 457 M
57 88 V
56 85 V
56 81 V
57 75 V
56 69 V
56 62 V
57 57 V
56 50 V
56 46 V
57 40 V
56 35 V
56 31 V
57 27 V
56 24 V
56 20 V
57 17 V
56 15 V
56 12 V
57 11 V
56 8 V
56 7 V
57 5 V
56 5 V
57 3 V
56 2 V
56 1 V
57 0 V
56 0 V
56 0 V
57 -2 V
56 -2 V
56 -2 V
57 -2 V
56 -4 V
56 -3 V
57 -3 V
56 -4 V
56 -4 V
57 -4 V
56 -4 V
56 -5 V
57 -4 V
56 -5 V
56 -4 V
57 -5 V
56 -4 V
56 -5 V
57 -4 V
56 -5 V
LT1
3114 1846 M
180 0 V
656 473 M
57 77 V
56 73 V
56 69 V
57 63 V
56 58 V
56 53 V
57 48 V
56 43 V
56 39 V
57 34 V
56 31 V
56 27 V
57 24 V
56 21 V
56 18 V
57 16 V
56 14 V
56 12 V
57 10 V
56 9 V
56 7 V
57 6 V
56 6 V
57 3 V
56 4 V
56 2 V
57 2 V
56 1 V
56 0 V
57 0 V
56 0 V
56 -2 V
57 0 V
56 -3 V
56 -2 V
57 -2 V
56 -2 V
56 -2 V
57 -3 V
56 -4 V
56 -2 V
57 -4 V
56 -3 V
56 -4 V
57 -3 V
56 -4 V
56 -4 V
57 -3 V
56 -4 V
LT2
3114 1746 M
180 0 V
656 360 M
57 61 V
56 58 V
56 56 V
57 52 V
56 49 V
56 46 V
57 42 V
56 38 V
56 36 V
57 32 V
56 30 V
56 27 V
57 24 V
56 22 V
56 20 V
57 18 V
56 16 V
56 14 V
57 13 V
56 11 V
56 10 V
57 9 V
56 9 V
57 7 V
56 6 V
56 5 V
57 4 V
56 5 V
56 3 V
57 3 V
56 2 V
56 2 V
57 1 V
56 1 V
56 1 V
57 0 V
56 0 V
56 0 V
57 0 V
56 -1 V
56 -1 V
57 -1 V
56 -2 V
56 -1 V
57 -2 V
56 -1 V
56 -2 V
57 -2 V
56 -2 V
LT3
3114 1646 M
180 0 V
656 321 M
57 50 V
56 49 V
56 47 V
57 44 V
56 41 V
56 39 V
57 36 V
56 33 V
56 31 V
57 29 V
56 26 V
56 25 V
57 22 V
56 20 V
56 19 V
57 17 V
56 15 V
56 14 V
57 13 V
56 12 V
56 10 V
57 10 V
56 9 V
57 7 V
56 7 V
56 6 V
57 6 V
56 5 V
56 4 V
57 4 V
56 4 V
56 3 V
57 2 V
56 3 V
56 2 V
57 2 V
56 1 V
56 1 V
57 0 V
56 1 V
56 1 V
57 0 V
56 0 V
56 -1 V
57 0 V
56 -1 V
56 0 V
57 -1 V
56 -1 V
LT4
3114 1546 M
180 0 V
656 733 M
57 9 V
56 43 V
56 53 V
57 56 V
56 56 V
56 53 V
57 49 V
56 45 V
56 42 V
57 37 V
56 33 V
56 30 V
57 26 V
56 24 V
56 21 V
57 19 V
56 15 V
56 15 V
57 12 V
56 11 V
56 9 V
57 9 V
56 6 V
57 6 V
56 5 V
56 5 V
57 3 V
56 3 V
56 2 V
57 2 V
56 1 V
56 2 V
57 0 V
56 0 V
56 1 V
57 -1 V
56 0 V
56 0 V
57 0 V
56 -1 V
56 -1 V
57 0 V
56 -1 V
56 -1 V
57 -1 V
56 0 V
56 -1 V
57 -1 V
56 0 V
stroke
grestore
end
showpage
}
\put(3054,1546){\makebox(0,0)[r]{Rotational mass for $s=0.2$}}
\put(3054,1646){\makebox(0,0)[r]{$s=0.4$}}
\put(3054,1746){\makebox(0,0)[r]{$s=0.3$}}
\put(3054,1846){\makebox(0,0)[r]{$s=0.2$}}
\put(3054,1946){\makebox(0,0)[r]{$s=0.13$}}
\put(100,1180){%
\special{ps: gsave currentpoint currentpoint translate
270 rotate neg exch neg exch translate}%
\makebox(0,0)[b]{\shortstack{$M/M_{\odot}$}}%
\special{ps: currentpoint grestore moveto}%
}
\put(3417,151){\makebox(0,0){1.8}}
\put(3041,151){\makebox(0,0){1.6}}
\put(2666,151){\makebox(0,0){1.4}}
\put(2290,151){\makebox(0,0){1.2}}
\put(1915,151){\makebox(0,0){1}}
\put(1539,151){\makebox(0,0){0.8}}
\put(1163,151){\makebox(0,0){0.6}}
\put(788,151){\makebox(0,0){0.4}}
\put(540,1971){\makebox(0,0)[r]{3}}
\put(540,1627){\makebox(0,0)[r]{2.5}}
\put(540,1283){\makebox(0,0)[r]{2}}
\put(540,939){\makebox(0,0)[r]{1.5}}
\put(540,595){\makebox(0,0)[r]{1}}
\put(540,251){\makebox(0,0)[r]{0.5}}
\end{picture}
\setlength{\unitlength}{0.1bp}
\special{!
/gnudict 40 dict def
gnudict begin
/Color false def
/Solid false def
/gnulinewidth 5.000 def
/vshift -33 def
/dl {10 mul} def
/hpt 31.5 def
/vpt 31.5 def
/M {moveto} bind def
/L {lineto} bind def
/R {rmoveto} bind def
/V {rlineto} bind def
/vpt2 vpt 2 mul def
/hpt2 hpt 2 mul def
/Lshow { currentpoint stroke M
  0 vshift R show } def
/Rshow { currentpoint stroke M
  dup stringwidth pop neg vshift R show } def
/Cshow { currentpoint stroke M
  dup stringwidth pop -2 div vshift R show } def
/DL { Color {setrgbcolor Solid {pop []} if 0 setdash }
 {pop pop pop Solid {pop []} if 0 setdash} ifelse } def
/BL { stroke gnulinewidth 2 mul setlinewidth } def
/AL { stroke gnulinewidth 2 div setlinewidth } def
/PL { stroke gnulinewidth setlinewidth } def
/LTb { BL [] 0 0 0 DL } def
/LTa { AL [1 dl 2 dl] 0 setdash 0 0 0 setrgbcolor } def
/LT0 { PL [] 0 1 0 DL } def
/LT1 { PL [4 dl 2 dl] 0 0 1 DL } def
/LT2 { PL [2 dl 3 dl] 1 0 0 DL } def
/LT3 { PL [1 dl 1.5 dl] 1 0 1 DL } def
/LT4 { PL [5 dl 2 dl 1 dl 2 dl] 0 1 1 DL } def
/LT5 { PL [4 dl 3 dl 1 dl 3 dl] 1 1 0 DL } def
/LT6 { PL [2 dl 2 dl 2 dl 4 dl] 0 0 0 DL } def
/LT7 { PL [2 dl 2 dl 2 dl 2 dl 2 dl 4 dl] 1 0.3 0 DL } def
/LT8 { PL [2 dl 2 dl 2 dl 2 dl 2 dl 2 dl 2 dl 4 dl] 0.5 0.5 0.5 DL } def
/P { stroke [] 0 setdash
  currentlinewidth 2 div sub M
  0 currentlinewidth V stroke } def
/D { stroke [] 0 setdash 2 copy vpt add M
  hpt neg vpt neg V hpt vpt neg V
  hpt vpt V hpt neg vpt V closepath stroke
  P } def
/A { stroke [] 0 setdash vpt sub M 0 vpt2 V
  currentpoint stroke M
  hpt neg vpt neg R hpt2 0 V stroke
  } def
/B { stroke [] 0 setdash 2 copy exch hpt sub exch vpt add M
  0 vpt2 neg V hpt2 0 V 0 vpt2 V
  hpt2 neg 0 V closepath stroke
  P } def
/C { stroke [] 0 setdash exch hpt sub exch vpt add M
  hpt2 vpt2 neg V currentpoint stroke M
  hpt2 neg 0 R hpt2 vpt2 V stroke } def
/T { stroke [] 0 setdash 2 copy vpt 1.12 mul add M
  hpt neg vpt -1.62 mul V
  hpt 2 mul 0 V
  hpt neg vpt 1.62 mul V closepath stroke
  P  } def
/S { 2 copy A C} def
end
}
\begin{picture}(3600,2160)(0,0)
\special{"
gnudict begin
gsave
50 50 translate
0.100 0.100 scale
0 setgray
/Helvetica findfont 100 scalefont setfont
newpath
-500.000000 -500.000000 translate
LTa
LTb
600 251 M
63 0 V
2754 0 R
-63 0 V
600 623 M
63 0 V
2754 0 R
-63 0 V
600 994 M
63 0 V
2754 0 R
-63 0 V
600 1366 M
63 0 V
2754 0 R
-63 0 V
600 1737 M
63 0 V
2754 0 R
-63 0 V
600 2109 M
63 0 V
2754 0 R
-63 0 V
788 251 M
0 63 V
0 1795 R
0 -63 V
1163 251 M
0 63 V
0 1795 R
0 -63 V
1539 251 M
0 63 V
0 1795 R
0 -63 V
1915 251 M
0 63 V
0 1795 R
0 -63 V
2290 251 M
0 63 V
0 1795 R
0 -63 V
2666 251 M
0 63 V
0 1795 R
0 -63 V
3041 251 M
0 63 V
0 1795 R
0 -63 V
3417 251 M
0 63 V
0 1795 R
0 -63 V
600 251 M
2817 0 V
0 1858 V
-2817 0 V
600 251 L
LT0
3114 1946 M
180 0 V
656 519 M
57 79 V
56 73 V
56 66 V
57 60 V
56 53 V
56 48 V
57 42 V
56 37 V
56 33 V
57 28 V
56 25 V
56 22 V
57 18 V
56 17 V
56 14 V
57 12 V
56 10 V
56 8 V
57 8 V
56 6 V
56 5 V
57 4 V
56 3 V
57 3 V
56 2 V
56 1 V
57 0 V
56 1 V
56 -1 V
57 0 V
56 0 V
56 -1 V
57 -2 V
56 -1 V
56 -1 V
57 -1 V
56 -2 V
56 -1 V
57 -2 V
56 -1 V
56 -2 V
57 -1 V
56 -2 V
56 0 V
57 -2 V
56 -1 V
56 -1 V
57 -2 V
56 -1 V
LT1
3114 1846 M
180 0 V
656 466 M
57 94 V
56 89 V
56 81 V
57 75 V
56 66 V
56 59 V
57 50 V
56 45 V
56 38 V
57 32 V
56 29 V
56 23 V
57 20 V
56 17 V
56 14 V
57 12 V
56 9 V
56 8 V
57 6 V
56 5 V
56 4 V
57 3 V
56 2 V
57 1 V
56 1 V
56 1 V
57 0 V
56 -1 V
56 -1 V
57 -1 V
56 -2 V
56 -1 V
57 -2 V
56 -2 V
56 -2 V
57 -2 V
56 -2 V
56 -3 V
57 -2 V
56 -2 V
56 -2 V
57 -2 V
56 -2 V
56 -2 V
57 -1 V
56 -2 V
56 -1 V
57 -2 V
56 -1 V
LT2
3114 1746 M
180 0 V
656 491 M
769 653 L
882 795 L
994 915 L
113 99 V
113 78 V
112 63 V
113 48 V
113 37 V
112 28 V
113 17 V
113 0 V
113 0 V
112 0 V
113 0 V
113 0 V
112 -2 V
113 -7 V
113 -9 V
112 -10 V
113 -9 V
113 -10 V
112 -10 V
113 -10 V
113 -10 V
LT3
3114 1646 M
180 0 V
656 491 M
769 653 L
882 795 L
994 915 L
113 99 V
113 78 V
112 63 V
113 48 V
113 37 V
112 28 V
113 21 V
113 14 V
113 10 V
112 0 V
113 0 V
113 0 V
112 0 V
113 0 V
113 0 V
112 0 V
113 0 V
113 -2 V
112 -6 V
113 -8 V
113 -9 V
LT4
3114 1546 M
180 0 V
656 744 M
57 12 V
56 56 V
56 68 V
57 70 V
56 65 V
56 60 V
57 54 V
56 47 V
56 41 V
57 36 V
56 30 V
56 26 V
57 23 V
56 19 V
56 16 V
57 13 V
56 11 V
56 10 V
57 7 V
56 6 V
56 5 V
57 4 V
56 3 V
57 2 V
56 2 V
56 0 V
57 1 V
56 0 V
56 0 V
57 -1 V
56 0 V
56 -2 V
57 -1 V
56 -2 V
56 -1 V
57 -2 V
56 -2 V
56 -1 V
57 -3 V
56 -1 V
56 -2 V
57 -2 V
56 -1 V
56 -2 V
57 -1 V
56 -1 V
56 -1 V
57 -1 V
56 -1 V
stroke
grestore
end
showpage
}
\put(3054,1546){\makebox(0,0)[r]{Rotational mass for $B=200$ MeVfm$^{-3}$ }}
\put(3054,1646){\makebox(0,0)[r]{Maxwell $B=200$ MeVfm$^{-3}$ }}
\put(3054,1746){\makebox(0,0)[r]{Maxwell $B=150$ MeVfm$^{-3}$ }}
\put(3054,1846){\makebox(0,0)[r]{Mixed phase $B=200$ MeVfm$^{-3}$ }}
\put(3054,1946){\makebox(0,0)[r]{Mixed phase $B=150$ MeVfm$^{-3}$ }}
\put(2008,20){\makebox(0,0){$n_c$ (fm$^{-3}$)}}
\put(100,1180){%
\special{ps: gsave currentpoint currentpoint translate
270 rotate neg exch neg exch translate}%
\makebox(0,0)[b]{\shortstack{$M/M_{\odot}$}}%
\special{ps: currentpoint grestore moveto}%
}
\put(3417,151){\makebox(0,0){1.8}}
\put(3041,151){\makebox(0,0){1.6}}
\put(2666,151){\makebox(0,0){1.4}}
\put(2290,151){\makebox(0,0){1.2}}
\put(1915,151){\makebox(0,0){1}}
\put(1539,151){\makebox(0,0){0.8}}
\put(1163,151){\makebox(0,0){0.6}}
\put(788,151){\makebox(0,0){0.4}}
\put(540,2109){\makebox(0,0)[r]{3}}
\put(540,1737){\makebox(0,0)[r]{2.5}}
\put(540,1366){\makebox(0,0)[r]{2}}
\put(540,994){\makebox(0,0)[r]{1.5}}
\put(540,623){\makebox(0,0)[r]{1}}
\put(540,251){\makebox(0,0)[r]{0.5}}
\end{picture}
\caption{Total mass $M$ as function of central density $n_c$
         for various values of $s$ (upper panel) 
         and the bag parameter $B$ (lower panel) for
         both a mixed phase and a Maxwell contructed EoS with 
         $s=0.2$ in Eq.~\ref{eq:EA}). 
         In addition we include also the rotational corrections
         for the pure $pn$-case with $s=0.2$ and the mixed
         phase contruction for $B=200$ MeVfm$^{-3}$. For the Maxwell
         construction which exhibits a first order phase transition, 
         in the density regions where the two phases coexist, 
         the pressure is constant, a fact reflected
         in the constant value of the neutron star mass. All results
         are for $\beta$-stable matter. Note also that for the upper
         panel, the EoS for $s=0.3$ and $s=0.4$ start to differ
        from those with $s=0.13,0.2$ at densities below $0.2$ fm$^{-3}$.}
\label{fig2}
\end{figure}
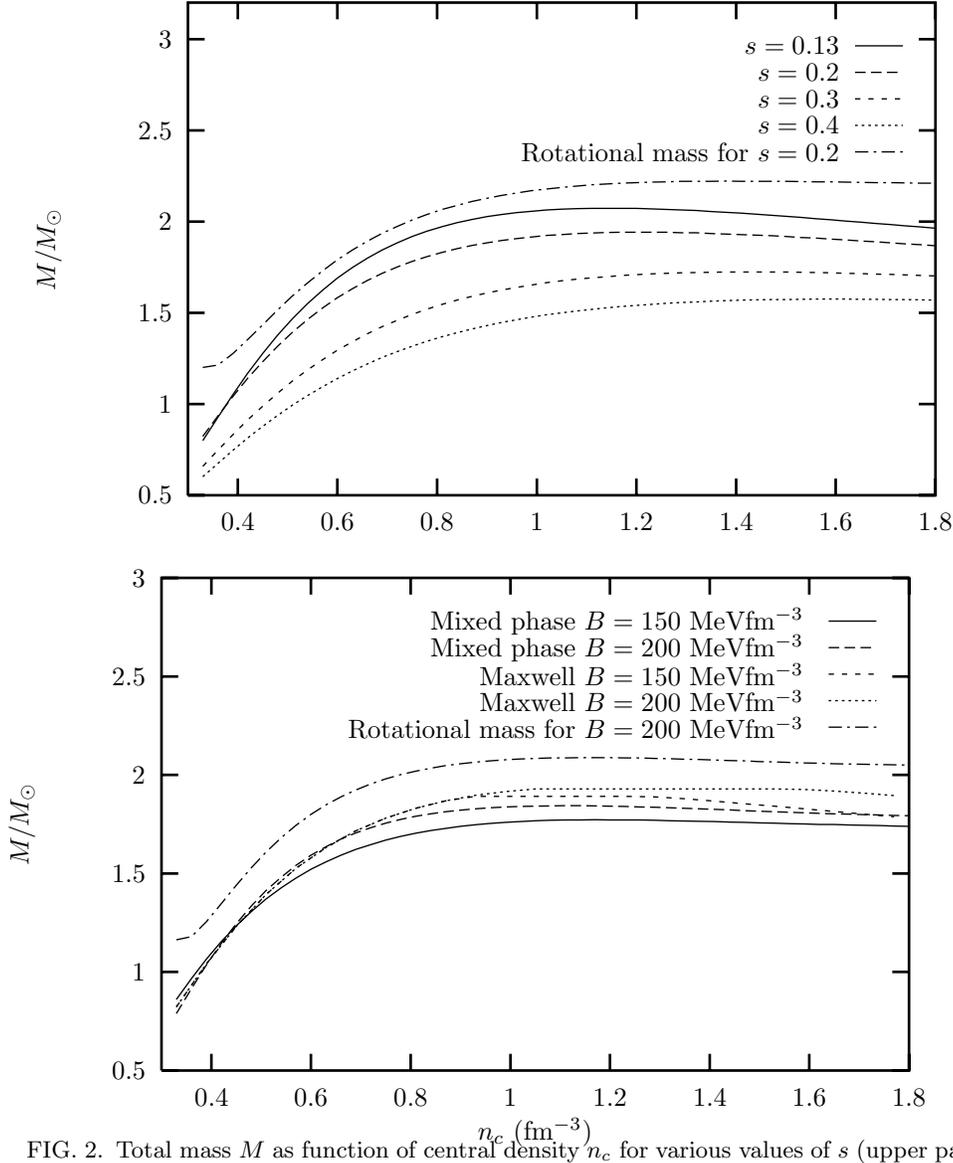

As can be seen from Fig.\ \ref{fig2} none of the equations of state
from either the pure $pn$ phase or with a mixed phase or Maxwell
construction with quark degrees of freedom, result in stable
configurations for densities above $\sim 10 n_0$, implying thereby
that none of the stars have cores with a pure quark phase.  The EoS
with $pn$ degrees of freedom have masses $M\la2.2M_{\odot}$ when
rotational corrections are accounted for.  With the inclusion of the
mixed phase, the total mass is reduced since the EoS is softer.
For pure
quark stars there is only one energy scale namely $B$ which provides
a homology transformation (Madsen 1998) and the maximum mass is
$M_{max}=2.0M_\odot (58{\rm MeV fm^{-3}}/B)^{1/2}$ (for
$\alpha_s=0$). However, for $B\ga 58{\rm MeV fm^{-3}}$ a nuclear
matter mantle has to be added and for $B\la 58{\rm MeV fm^{-3}}$ quark
matter has lower energy per baryon than $^{56}$Fe and is thus the
ground state of strongly interacting matter. Unless the latter is the
case, we can thus exclude the existence of $2.2-2.3M_\odot$ quark
stars.

\begin{figure}
   \setlength{\unitlength}{1mm}
   \begin{picture}(100,140)
   \put(25,0){\epsfxsize=14cm \epsfbox{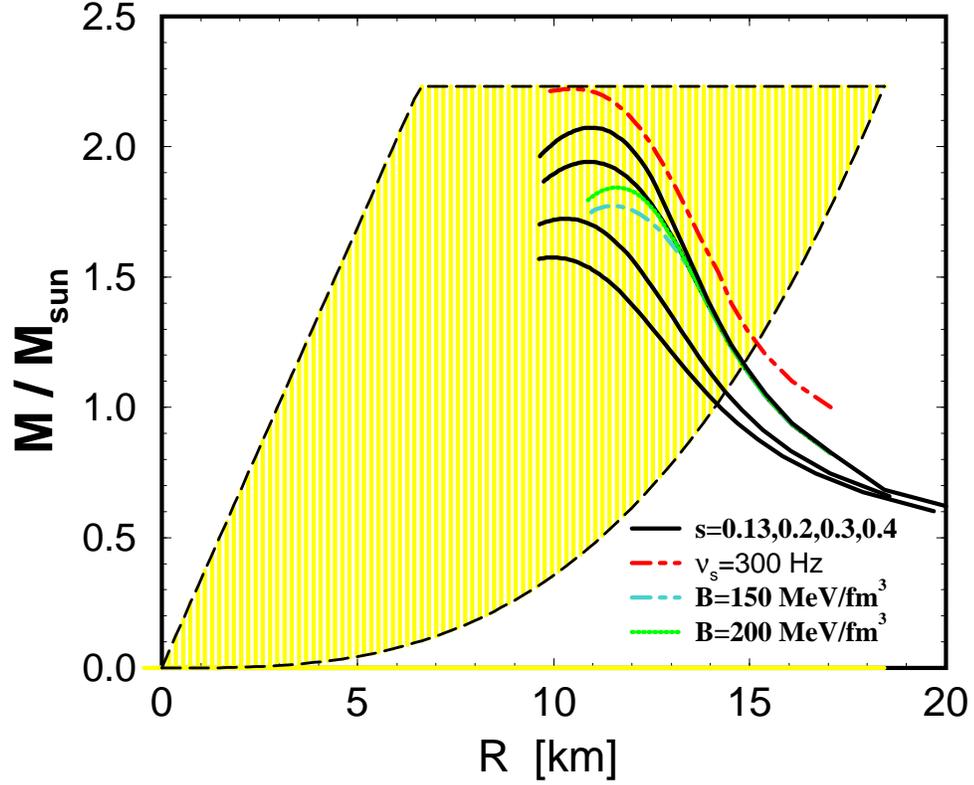}}
   \end{picture}
\caption{Neutron star masses vs.~radius for the EoS of Eq.~(1) 
with softness s=0.13,0.2,0.3,04, with increasing values of $s$ 
from top to bottom for the full curves.
Phase transitions decrease the maximum mass whereas rotation
increases it. The shaded area represents  the neutron star
radii and masses allowed (see text) for 
orbital QPO frequencies 1060~Hz of 4U 1820-30.
\label{fig3}}
\end{figure}

In Fig.\ \ref{fig3} we show the mass-radius relations for the various
equations of state. 
The shaded area represents the allowed masses and radii for 
$\nu_{QPO}=1060$ Hz of 4U 1820-30. Generally,
\begin{eqnarray}
  2GM < R < \left(\frac{GM}{4\pi^2\nu_{QPO}^2}\right)^{1/3} \,,
\end{eqnarray}
where the lower limit ensures that the star is not a black hole,
and the upper limit that the accreting matter orbits outside
the star, $R<R_{orb}$. Furthermore,
for the matter to be outside the innermost
stable orbit, $R>R_{ms}=6GM$, requires that 
\begin{eqnarray}
   M &\la& \frac{1+0.75j}{12\sqrt{6}\pi G\nu_{QPO}}  \label{Mms} \\
 &\simeq& 2.2 M_\odot (1+0.75j)\frac{{\rm kHz}}{\nu_{QPO}} \,, \nonumber
\end{eqnarray}
where $j=2\pi c\nu_sI/M^2$ is a dimensionless measure of the angular
momentum of the star with moment of inertia $I$.  The upper limit
in Eq. (\ref{Mms}) is the mass when $\nu_{QPO}$ corresponds to the innermost
stable orbit. According to Zhang, Smale, Strohmayer \& Swank (1998)
this is the case for 4U 1820-30 since
$\nu_{QPO}$ saturates at $\sim1060$~Hz with increasing count rate.
The corresponding neutron star mass is $M\sim 2.2-2.3M_\odot$ which
leads to several interesting conclusions as seen in Fig.\
\ref{fig3}. Firstly, the stiffest EoS allowed by causality
($s\simeq 0.13-0.2$) is needed. Secondly, rotation must be included
which increases the maximum mass and corresponding
radii by 10-15\% for $\nu_s\sim 300$~Hz.
Thirdly, a phase transition to quark matter below densities of order
$\sim 5 n_0$ can be excluded, corresponding to a restriction on the bag
constant $B\ga200$ MeVfm$^{-3}$.

 These maximum masses are smaller than those of APR98 and Kalogera \&
Baym (1996) who, as discussed above, obtain upper bounds on the mass
of neutron stars by discontinuously setting the sound speed to equal
the speed of light above a certain density, $n_c$. By varying the density
$n_c=2\to 5n_0$ the
maximum mass drops from $2.9\to 2.2M_\odot$. In our case,
incorporating causality smoothly by introducting the parameter $s$ in
Eq.\ (\ref{eq:EA}), the EoS is softened at higher densities in order
to obey causality, and yields a maximum mass which instead is slightly
lower than the $2.2M_\odot$ derived in APR98 for nonrotating stars.

If the QPOs are not from the innermost stable orbits and one finds
that even accreting neutron stars have small masses, say like the
binary pulsars, $M\la1.4M_\odot$, this may indicate that heavier
neutron stars are not stable. Therefore, the EoS is soft at high
densities $s\ga0.4$ or that a phase transition occurs at few
times nuclear matter densities. For the nuclear to quark matter
transition this would require $B<80$ MeVfm$^{-3}$ for
$s=0.2$. For such small bag parameters there is an appreciable
quark and nuclear matter mixed phase in the neutron star interior but
even in these extreme cases a pure quark matter core is not obtained
for stable neutron star configurations.

A third QPO frequency referred to as Horisontal Branch Oscillations
around $\nu_{HBO}\simeq 20-50$ Hz has been suggested to be
caused by Lense-Thirring precession at the inner border of the
accretion disk (Stella \& Vietri 1998)
\begin{eqnarray}
  \nu_{LT} &=& \frac{8\pi^2 I}{c^2M} \nu_s\nu^2_{QPO}  \nonumber\\
  &\simeq& \frac{13.2}{50{\rm km^2}}\frac{I}{M}
   \frac{\nu_s}{300{\rm H}z}\left(\frac{\nu_{QPO}}{{\rm 1kHz}}\right)^2
  \,. \label{LT}
\end{eqnarray}
However, 
even for the stiffest EoS $s\simeq 0.13-0.2$ 
we calculate moment of inertia and Lense-Thirring
frequencies from  Eq.~(\ref{LT}), which are 
a factor $\sim4$ below the observed $\nu_{HBO}$ thus
confirming analyses of Schaab \& Weigel (1999), Psaltis et al.~(1999) and
Kalogera \& Psaltis (1999).

\section{SUMMARY}

Modern nucleon-nucleon potentials have reduced the uncertainties in
the calculated EoS.  Using the most recent realistic effective
interactions for nuclear matter of APR98 with a smooth extrapolation
to high densities including causality, the equation of state could be
constrained by a ``softness'' parameter $s$ which parametrizes
the unknown stiffness of the EoS at high densities. Maximum masses
were calculated for rotating neutron stars with and without first and
second order phase transitions to, e.g., quark matter at high
densities.

The calculated bounds for maximum masses leaves two natural options
when compared to the observed neutron star masses:
\begin{itemize}
 \item {\bf Case I}: {\it The large masses of the neutron stars in
QPO 4U 1820-30 ($M=2.3M_\odot$), PSR J1012+5307
($M=2.1\pm0.4 M_\odot$), Vela X-1 ($M=1.9\pm0.1 M_\odot$), and
Cygnus X-2 ($M=1.8\pm0.4 M_\odot$), are confirmed and
complemented by other neutron stars with masses around $\sim 2M_\odot$.}

As a consequence, the EoS of dense nuclear matter is severely
restricted and only the stiffest EoS consistent with causality are allowed,
i.e., softness parameter $0.13\le s\la0.2$.  Furthermore, any
significant phase transition at densities below $< 5n_0$ can be
excluded. 
 That the radio binary pulsars all have masses around $1.4M_\odot$ is
then probably due to the formation mechanism in supernovae where the
Chandrasekhar mass for iron cores are $\sim1.5M_\odot$.
Neutron stars in binaries can subsequently acquire larger
masses by accretion as X-ray binaries.

 \item {\bf Case II}: 
{\it The heavy neutron stars prove erroneous by more detailed observations
and only masses like those of binary pulsars are found.}
If accretion does not produce neutron stars heavier than 
$\ga1.4M_\odot$, this indicates that heavier neutron stars simply are not
stable which in turn implies a soft EoS, either $s> 0.4$ or a
significant phase transition must occur already at a few times nuclear
saturation densities. 
\end{itemize}

\end{multicols}

\end{document}